\documentclass[prd,preprint,showpacs]{revtex4}
\usepackage{graphicx}
\newcommand{\be}{\begin{equation}}
\newcommand{\ee}{\end{equation}}
\newcommand{\bea}{\begin{eqnarray}}
\newcommand{\eea}{\end{eqnarray}}
\newcommand{\bean}{\begin{eqnarray*}}
\newcommand{\eean}{\end{eqnarray*}}
\begin{document}
%
\title{Studying $K\pi~S$-wave scattering in K-matrix formalism}
\date{November 3, 2002}
\author{Long Li}
\email{lilong@mail.ihep.ac.cn}
\affiliation{Institute of High Energy Physics,
P.O. Box 918(4), Beijing 100039, China}
\author{Bing-song Zou}
\email{zoubs@mail.ihep.ac.cn}
\affiliation{{CCAST} (World Laboratory),
P.O. Box 8730, Beijing 100080, China\\
Institute of High Energy Physics,
P.O. Box 918(4), Beijing 100039, China\\
and Institute of Theoretical Physics,
Beijing 100080, China}
\author{Guang-lie Li}
\email{ligl@mail.ihep.ac.cn}
\affiliation{{CCAST} (World Laboratory),
P.O. Box 8730, Beijing 100080, China\\
Institute of High Energy Physics,
P.O. Box 918(4), Beijing 100039, China\\
and Center of Theoretical Nuclear Physics,
National Laboratory of Heavy Ion accelerator,
Lanzhou 730000, China}
%
%
\begin{abstract}
We generalize our previous work on $\pi\pi$ scattering to $K\pi$
scattering, and re-analyze the experiment data of $K\pi$
scattering below 1.6 GeV. Without any free parameter, we explain
$K\pi~I=3/2~S$-wave phase shift very well by using $t$-channel
$\rho$ and $u$-channel $K^*$ meson exchange.  With the $t$-channel
and $u$-channel meson exchange fixed as the background term, we
fit the $K\pi~I=1/2~S$-wave data of the LASS experiment quite well
by introducing one or two $s$-channel resonances. It is found that
there is only one $s$-channel resonance between $K\pi$ threshold
and 1.6 GeV, {\sl i.e.}, $K_0^*(1430)$ with a mass around
$1438\sim 1486$ MeV and a width about 346 MeV, while the
$t$-channel $\rho$ exchange gives a pole at ($450-480i$) MeV for
the amplitude.
\end{abstract}
\maketitle
%
%
\section{introduction}
The assignment of the scalar mesons has been a long standing
problem. Recently the existence of the low-lying $\pi\pi$ scalar
state $\sigma$ has been well established,  \textit{i.e.},
$f_0(400$-$1200)$ as listed by the Particle Data Group (PDG)
\cite{PDG2000}. Now the PDG lists five well-established isoscalar
$0^{++}$ mesons: $f_0$(400-1200), $f_0(980)$, $f_0(1370)$,
$f_0(1500)$ and $f_0(1710)$, which are obviously too many for a
standard $q\bar q$ nonet. Given the existence of two isovector
scalars, $a_0(980)$ and $a_0(1450)$, two scalar nonets have been
suggested \cite{Beveren1999:epjc10,Black2000:prd61}: an unconventional one
composed of $\sigma$, $\kappa$, $a_0(980)$ and $f_0(980)$ and the
conventional $q\bar q$ nonet composed of $f_0(1370)$,
$K^*_0(1430)$, $a_0(1450)$ and $f_0(1500)$ or $f_0(1710)$.

However, the existence of the $\kappa$ is still in controversy.
Evidence for this resonance has been claimed within certain models
\cite{Beveren1986:zpc30,Ishida1997:ptp98,Black1998:prd58,Oller1999:prd60},
whilst other studies
dispute this \cite{Tornqvist1995:zpc68,Anisovich1997:plb413}. Recently, a
less model-dependent analysis of the LASS $K\pi$ scattering data
between 825 MeV and 2 GeV by Cherry and Pennington
\cite{Cherry2001:npa688} concludes that there is no $\kappa(900)$,
but a very low mass $\kappa$ well below 825 MeV cannot be ruled
out.

In fact the phase shifts of $K\pi~S$-wave scattering at low
energies \cite{Aston1988:npb296,Estabrooks1978:npb133} look very
similar to those of $\pi\pi~S$-wave scattering. In our previous
study on $\pi\pi$ scattering in the K-matrix formalism
\cite{LongLi2001:prd63}, the negative phase shifts for the
isotensor $\pi\pi~S$-wave were naturally explained by the
$t$-channel $\rho$ meson exchange while the broad
$f_0(400$-$1200)$ structure in the isoscalar $\pi\pi~S$-wave was
decomposed into a $t$-channel $\rho$ meson exchange part
dominating at the low energy end plus an additional $s$-channel
wide resonance $f_0(1670)$. Considering the similarity between the
$K\pi$ scattering and the $\pi\pi$ scattering, it is nature to
extend our previous work on the $\pi\pi$ scattering to the $K\pi$
scattering. We find that the negative phase shifts of the
$K\pi~I=3/2~S$-wave can be very well reproduced by the $t$-channel
$\rho$ and $u$-channel $K^*$ meson exchange without any free
parameter. With the $t$-channel and $u$-channel meson exchange
fixed as the background term, the positive smoothly rising phase
shifts for the $K\pi~I=1/2~S$-wave can be well fitted by
introducing one or two additional $s$-channel resonances. It is
found that there is only one $s$-channel resonance between $K\pi$
threshold and 1.6 GeV, {\sl i.e.}, $K_0^*(1430)$ with a mass
around $1438\sim 1486$ MeV and a width about 346 MeV, while the
$t$-channel $\rho$ exchange gives a pole at ($450-480i$) MeV for
the amplitude.
\section{formalism}
For the pseudoscalar-pseudoscalar-vector coupling, we use the
SU(3)-symmetric lagrangian \cite{Lohse1990:npa516} \be
\mathcal{L}_{PPV}=-\frac{1}{2}iG_V\text{Tr}([P,\partial_\mu
P]V^\mu), \ee where $G_V$ is the coupling constant, $P$ is the
$3\times3$ matrix representation of the pseudoscalar meson octet,
$P=\lambda^aP^a, a=1,\ldots,8$ and $\lambda^a$ are the $3\times3$
generators of SU(3).  A similar definition of $V^\mu$ is used for
the vector meson octet.

In the Gell-Mann representation, the lagrangian can be expressed
as \be \mathcal{L}_{PPV}=2G_Vf_{abc}P^a\partial_\mu P^bV^{c\mu},
\ee where $f_{abc}$ are the antisymmetric structure constants of
SU(3). For example, \be
\mathcal{L}_{\pi\pi\rho}=2G_V\epsilon_{ijk}\pi^i\partial_\mu\pi^j\rho^{k\mu},
\ee and \bea \mathcal{L}_{\pi KK^*}&=&iG_V\left\{
\left[(\partial_\mu \overline{K})\vec{\tau}K^{*\mu}
-\overline{K^*}^\mu\vec{\tau}(\partial_\mu K)\right]\cdot
\vec{\pi} -\left[\overline{K}\vec{\tau}K^{*\mu}
-\overline{K^*}^\mu\vec{\tau}K\right]\cdot (\partial_\mu
\vec{\pi}) \right\} \eea where
$\vec{\pi}\equiv\left(\pi_1,\pi_2,\pi_3\right)$,
$K^{*\mu}\equiv\left(\begin{array}{c}K^{*+\mu}\\K^{*0\mu}\end{array}\right)$,
$K\equiv\left(\begin{array}{c}K^{+}\\K^{0}\end{array}\right)$,
$\overline{{K^*}}^\mu\equiv\left(K^{*-\mu},\overline{K^{*0}}^\mu\right)$,
$\overline K\equiv\left(K^{-},\overline{K^{0}}\right)$ and
$\vec{\tau}=(\tau_1,\tau_2,\tau_3)$ are usual Pauli matrices
acting on the kaon iso-spinors.

For $K\pi$ scattering, amplitude $T$ can be written in terms of
two invariant amplitudes $T^+$ and $T^-$ by \cite{Martin1974} \be
T_{\beta\alpha}=\delta_{\beta\alpha}T^++
\frac{1}{2}[\tau_\beta,\tau_\alpha]T^-, \ee where $\alpha,\beta$
are the isospin indices of the pions. Using isospin projection
operators gives \be
\begin{array}{l}
3T^+=T^{1/2}(s,t,u)+2T^{3/2}(s,t,u),\\
3T^-=T^{1/2}(s,t,u)-T^{3/2}(s,t,u).
\end{array}
\ee
where $s,t,u$ are the usual Mandelstam variables.

The partial-wave amplitudes are obtained from the full
amplitude by the standard projection formula
\cite{LongLi2001:prd63,Martin1974}
\bea
T_l(s)&=&\frac{1}{16\pi}\frac{1}{2}\int^{+1}_{-1}d(\cos\theta)
P_{l}(\cos\theta)T(s,t,u)
\nonumber\\
&=&\frac{1}{16\pi}\frac{1}{4p^2}\int^{0}_{-4p^2}dtP_l\left[1+\frac{t}{2p^2}\right]
T(s,t,u),
\eea
where $P_l(x)$ is the Legendre function and
$p=\sqrt{[s-(m_\pi+m_K)^2][s-(m_\pi-m_K)^2]}/(2\sqrt{s})$.
Our normalization is such that the unitarity relation for
partial-wave amplitude reads
$$\mathrm{Im}T_l(s)=\rho_1(s)|T_l(s)|^2,$$
with $\rho_1(s)=2p/\sqrt{s}$.

\begin{figure}[h]
\includegraphics{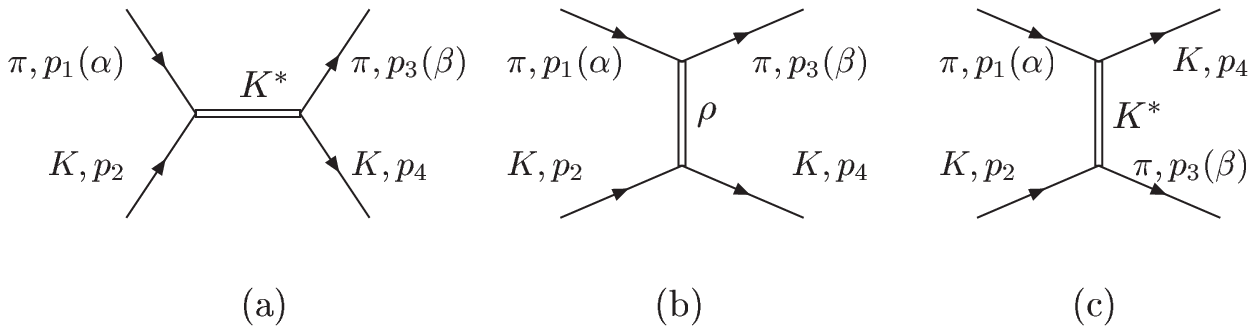}
\caption{\label{fig:PiKBorn}
The Born term of $K\pi$ scattering.}
\end{figure}

We start with the Born term of the $K\pi$ scattering amplitude
by $\rho$ meson and $K^*$meson exchange and follow the $K$-matrix
formalism as in
Refs.\cite{LongLi2001:prd63,Zou1994:prd50,Locher1997:prd55}.
Fig.\ref{fig:PiKBorn}
is the Feynman diagram of the $K\pi$ scattering Born term.
%

\subsection{$s$-channel and $u$-channel $K^*$ meson exchange amplitude}

The Born term for the $K^*$ meson exchange ((a) and (c) of
Fig.\ref{fig:PiKBorn}) is
\bea
T^{1/2}(s,t,u)&=&
g^2_{\pi KK^*}\left[\frac{3(t-u)}{m^2_{K^*}-s}+\frac{3(m^2_\pi-m^2_K)^2}{s(m^2_{K^*}-s)}
+\frac{s-t}{m^2_{K^*}-u}-\frac{(m^2_\pi-m^2_K)^2}{m^2_{K^*}(m^2_{K^*}-u)}\right],\\
T^{3/2}(s,t,u)&=&
-2g^2_{\pi KK^*}\left[\frac{s-t}{m^2_{K^*}-u}-\frac{(m^2_\pi-m^2_K)^2}{m^2_{K^*}(m^2_{K^*}-u)}\right],
\eea
where $g_{\pi KK^*}=G_V$ is the coupling constant.
Their $S$-wave projections are
\bea
K^{1/2}_S(s)&=&-\frac{1}{2}K^{3/2}_S(s)
\nonumber\\
&=&G_2\Bigg\{-1+
\frac{2(s-m^2_\pi-m^2_K)+m^2_{K^*}-(m^2_\pi-m^2_K)^2/m^2_{K^*}}{4p^2}\times
\nonumber\\
&&\ln\frac{m^2_{K^*}+s-2(m^2_\pi+m^2_K)}{m^2_{K^*}+s-2(m^2_\pi+m^2_K)-4p^2}\Bigg\},
\label{eq:kpiKstarI1/2BornS}
\eea
where $G_2={g^2_{\pi KK^*}}/(16\pi)$.
$K$-matrix unitarization is introduced by
\bea
T^{I=1/2}_S(s)&=&\frac{K^{I=1/2}_S(s)}{1-i\rho_1(s)K^{I=1/2}_S(s)},
\\
T^{I=3/2}_S(s)&=&\frac{K^{I=3/2}_S(s)}{1-i\rho_1(s)K^{I=3/2}_S(s)}.
\eea
%

Now we calculate the coupling constant $G_2$.  Considering
$I=1/2~P$-wave amplitude
\be
T^{I=1/2}_P(s)=\frac{K^{I=1/2}_P(s)}{1-i\rho_1(s)K^{I=1/2}_P(s)},
\ee
where $K^{I=1/2}_P$ is the $I=1/2~P$-wave Born amplitude
\bea
K^{1/2}_P(s)&=&
\frac{1}{4p^2}\int^{0}_{-4p^2}dt\Bigg\{G_2\Bigg[\frac{3(t-u)}{m^2_{K^*}-s}
+\frac{3(m^2_\pi-m^2_K)^2}{s(m^2_{K^*}-s)}+
\nonumber\\
&&\frac{s-t}{m^2_{K^*}-u}-
\frac{(m^2_\pi-m^2_K)^2}{m^2_{K^*}(m^2_{K^*}-u)}\Bigg]
\times\left[1+\frac{t}{2p^2}\right]\Bigg\}.
\eea
%

Near the $K^*$ pole at $s\approx m^2_{K^*}$, we have
\be
K^{1/2}_P(s)\approx \frac{G_24p^2}{m^2_{K^*}-s},
\ee
thus,
\be
T^{I=1/2}_P(s)=\frac{G_2 4p^2}{m^2_{K^*}-s-i\rho_1(s)G_2  4p^2}.
\ee
Comparing with the standard Breit-Wigner formula, we obtain
\be
M_{K^*}\Gamma_{K^*}=\rho_1(s)4p^2G_2|_{s=M_{K^*}},
\ee
which leads to $G_2=0.21$ with the $K^*$ mass $M_{K^*}=891.66$ MeV
and width $\Gamma_{K^*}=50.8$ MeV from Ref. \cite{PDG2000}.

The ratio of coupling constants is $g_{\rho\pi\pi}/g_{\pi
KK^*}\simeq 1.9$ using the $g_{\rho\pi\pi}$ value of
Refs.\cite{LongLi2001:prd63,Zou1994:prd50}:
$g^2_{\rho\pi\pi}/(32\pi)=0.364$. It agrees well with the value
from SU(3) symmetry: $g_{\rho\pi\pi}/g_{\pi KK^*}=2$.

In order to explain the $K\pi~I=3/2~S$-wave experimental
data, a form factor is needed to take into account the off-shell
behavior of the exchanged mesons.  For $t$ and $u$-channel exchange,
we use a form factor of conventional monopole type at each vertex:
\be
F(q)=\frac{\Lambda^2-m^2}{\Lambda^2-q^2},
\ee
where $m$ and $q$ are the mass and four-vector momentum of
exchanged mesons, and the cutoff parameter $\Lambda=1500$ MeV, the
same value as the $\pi\pi$ scattering in
Ref.\cite{LongLi2001:prd63}.

After adding the form factor, $K^{I=1/2}_S(s)$ and $K^{I=3/2}_S(s)$
becomes
\bea K^{1/2}_S(s)
&=&\frac{1}{4p^2}\int^{0}_{-4p^2}dt\Bigg\{G_2\left[
\frac{3(t-u)}{m^2_{K^*}-s}+\frac{3(m^2_\pi-m^2_K)^2}{s(m^2_{K^*}-s)}\right]
\nonumber\\
&&+\left(\frac{\Lambda^2-m^2_{K^*}}{\Lambda^2-u}\right)^2\left[\frac{s-t}{m^2_{K^*}-u}
-\frac{(m^2_\pi-m^2_K)^2}{m^2_{K^*}(m^2_{K^*}-u)}\right]
\Bigg\}
\nonumber\\
&=&G_2\Bigg\{\frac{m^2_{K^*}-\Lambda^2}{A-4p^2}\times
\left[1+\frac{s}{A}-\frac{(m^2_\pi-m^2_K)^2}{m^2_{K^*}A}\right]+
\nonumber\\
&&\frac{s+B-(m^2_\pi-m^2_K)^2/m^2_{K^*}}{4p^2}\ln\frac{B(A-4p^2)}{A(B-4p^2)}\Bigg\},
\\&&\nonumber\\
K^{3/2}_S(s)
&=&\frac{1}{4p^2}\int^{0}_{-4p^2}dt\left\{-2G_2\left(
\frac{\Lambda^2-m^2_{K^*}}{\Lambda^2-u}\right)^2
\left[\frac{s-t}{m^2_{K^*}-u}
-\frac{(m^2_\pi-m^2_K)^2}{m^2_{K^*}(m^2_{K^*}-u)}\right]\right\}
\nonumber\\
&=&-2G_2\Bigg\{\frac{m^2_{K^*}-\Lambda^2}{A-4p^2}\times
\left[1+\frac{s}{A}-\frac{(m^2_\pi-m^2_K)^2}{m^2_{K^*}A}\right]
\nonumber\\
&&+\frac{s+B-(m^2_\pi-m^2_K)^2/m^2_{K^*}}{4p^2}
\ln\frac{B(A-4p^2)}{A(B-4p^2)}\Bigg\},
\eea
where $A=\Lambda^2+s-2(m^2_\pi+m^2_K),B=m_{K^*}^2+s-2(m^2_\pi+m^2_K).$
%
\subsection{$t$-channel $\rho$ meson exchange amplitude}
The Born term for the $\rho$ meson exchange (see Fig.\ref{fig:PiKBorn} (b)) is
\bea
T^{Born}(I=1/2)&=&2g_{\pi\pi\rho}g_{\rho KK}\frac{s-u}{m^2_\rho-t},\\
T^{Born}(I=3/2)&=&-g_{\pi\pi\rho}g_{\rho KK}\frac{s-u}{m^2_\rho-t}.
\eea

 Their $S$-wave projections are
\bea
K^{1/2}_S(s)&=&-2K^{3/2}_S(s)
\nonumber\\
&=&2G_1\left\{-1+\frac{2(s-m^2_\pi-m^2_K)+m^2_\rho}{4p^2}
\ln\frac{m^2_\rho+4p^2}{m^2_\rho}\right\},
\eea
where $G_1=g^2_{\pi\pi\rho}/(32\pi)=0.364$
\cite{LongLi2001:prd63,Zou1994:prd50}. Because we cannot obtain
$g_{\rho KK}$ from experiment, SU(3) symmetry
$g_{\pi\pi\rho}=2g_{\rho KK}$ is used.

After introducing form factor,
\bea
K^{1/2}_S(s)&=&-2K^{3/2}_S(s)
\nonumber\\
&=&2G_1\Bigg\{\left[\frac{2(s-m^2_\pi-m^2_K)}{\Lambda^2}+1\right]\times
\frac{m^2_\rho-\Lambda^2}{\Lambda^2+4p^2}-
\nonumber\\
&&\frac{2(s-m^2_\pi-m^2_K)+m^2_\rho}{4p^2}
\ln\frac{m^2_\rho(\Lambda^2+4p^2)}{\Lambda^2(m^2_\rho+4p^2)}\Bigg\}.
\eea

\subsection{Amplitude of s-channel S-wave resonances}
The phase shift is known to be elastic below 1300 MeV.  The
threshold for the $K\eta^\prime$ channel is at 1453 MeV and the
$K\eta$ channel is only weakly coupled to the $K\pi$ channel
\cite{Aston1988:npb296,Tornqvist1995:zpc68}. Considering the
$K\pi$ and $K\eta^\prime$ channels, the explicit form is
\be T=\frac{M\Gamma_{K\pi}/\rho_1(M^2)}
{M^2-s-i\left[M\Gamma_{K\pi}\frac{\rho_1(s)}{\rho_1(M^2)}
+M\Gamma_{K\eta'}\frac{\rho_2(s)}{\rho_2(M^2)}\right]} \ee
where $\rho_2(s)=\sqrt{[s-(m_{\eta^\prime}+m_K)^2][s-(m_{\eta^\prime}-m_K)^2]}\,/s$
is the phase space factor of $K\eta^\prime$.

When fitting the experimental data, we first try introducing one
$s$-channel resonance and then try introducing two such
resonances.

\section{numerical results and discussion}

As in Refs.\cite{LongLi2001:prd63,Zou1994:prd50}, we use
Dalitz-Tuan method to combine various components given in the last
section to get the full partial wave amplitudes and corresponding
phase shifts.

\begin{figure}[htbp]
\begin{center}
\includegraphics[scale=0.7]{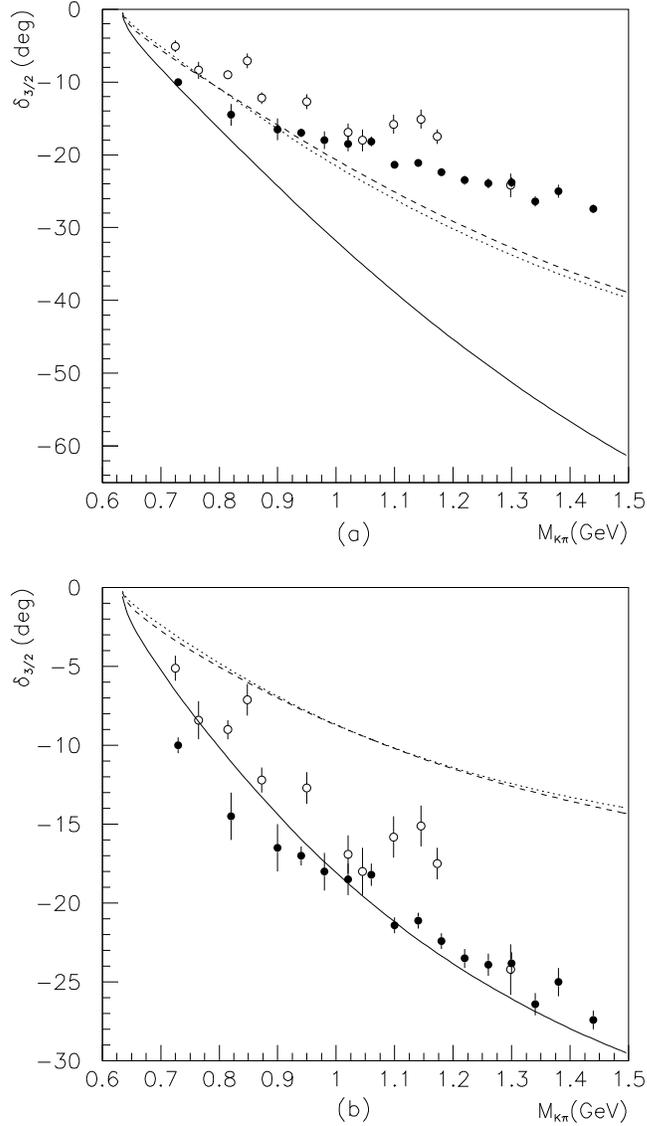}
\caption[$I=3/2~K\pi~S$-wave phase shift]
{\label{fig:kpiI3Phaseshift} $I=3/2~K\pi~S$-wave phase shift. Data
are from Refs.\cite{Estabrooks1978:npb133}(dots) and
\cite{Lang1978:fp26}(circles). Theoretical curves are for
$t$-channel $\rho$ exchange(dotted line), $u$-channel $K^*$
exchange (dashed line), and the sum (solid line). (a) without form
factor and (b) with form factor and $\Lambda=1.5$ GeV.
}
\end{center}
\end{figure}

For the $K\pi~I=3/2~S$-wave scattering, the phase shift is
negative with magnitude slowly increasing as the center-of-mass
energy increases as shown in Fig.\ref{fig:kpiI3Phaseshift}. There
is no s-channel quark-antiquark resonance contribution allowed for
isospin $I=3/2$. So the only contribution here is the $t$-channel
$\rho$ and $u$-channel $K^*$ meson exchanges. With the cutoff
parameter $\Lambda=1.5$ GeV fixed as the same as in $\pi\pi$
scattering \cite{LongLi2001:prd63}, we get the prediction for the
$K\pi~I=3/2~S$-wave phase shift as shown by the solid line in
Fig.\ref{fig:kpiI3Phaseshift}(b) without introducing any free
parameters, which reproduces data nicely. To show the effect of
off-shell form factor, the results without form factor are shown
in Fig.\ref{fig:kpiI3Phaseshift}(a). The $t$-channel $\rho$
exchange and $u$-channel $K^*$ exchange give very similar
contribution to the $K\pi~I=3/2~S$-wave phase shift as shown by
the dotted line and dashed line, respectively, in
Fig.\ref{fig:kpiI3Phaseshift}.

\begin{figure}[p]
\begin{center}
\includegraphics[scale=0.7]{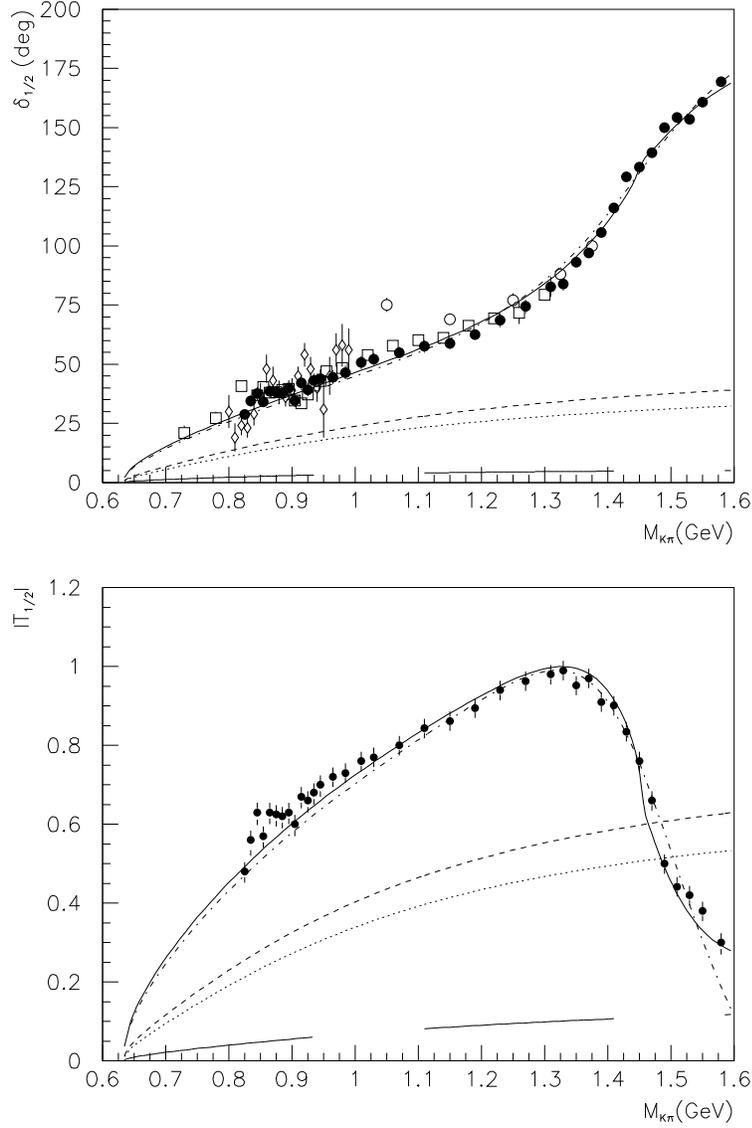}
\caption[The $I=1/2~K\pi~S$-wave phase shift and amplitude, only
including two resonances.]
{\label{fig:kpiI1PhaseshiftAndAmplitude} The $I=1/2~K\pi~S$-wave
phase shift and amplitude.  The experimental data for
$\delta_{1/2}$ and $T_{1/2}$ are from Ref.
\cite{Aston1988:npb296}(dotted), Ref.
\cite{Estabrooks1978:npb133}(boxed)
Ref. \cite{Firestone1972:prd5}(circled)
Ref. \cite{Matison1973:prd9}(diamond).
The long-dashed line is for $K^*$ meson exchange, the dotted line
is for $\rho$ meson exchange, the dashed line is the sum of $K^*$
and $\rho$ meson exchange, the dot-dashed line includes one
$s$-channel resonance and the solid line includes two resonances.
}
\end{center}
\end{figure}

Now we turn to the $K\pi~I=1/2~S$-wave scattering. The data and
our theoretical curves for the phase shift and amplitude magnitude
are shown in Fig.\ref{fig:kpiI1PhaseshiftAndAmplitude}. The
u-channel $K^*$ exchange and the t-channel $\rho$ meson exchange
with $\Lambda=1.5$ GeV give contributions as shown by the
long-dashed line and dotted line, respectively. Here the t-channel
$\rho$ exchange gives a much larger contribution than the
u-channel $K^*$ meson exchange. The sum of these two contributions
is shown by the dashed line and is obviously not enough to
reproduce the experimental data. Some contribution from s-channel
resonance(s) is definitely needed. By fixing the t-channel $\rho$
exchange and the u-channel $K^*$ exchange as background
contribution, we fit the LASS data \cite{Aston1988:npb296} first
by introducing one s-channel resonance (dot-dashed line) and then
by introducing two s-channel resonances (solid line). The fitted
parameters for the s-channel resonance(s) and the corresponding
$\chi^2$ for two cases are listed in Table
\ref{tab:kpiI1ResonancesParameters}.
\begin{table}[htb]
\begin{center}
\vspace{0.5cm}
\begin{tabular}{c|c|c|c|c|c|c}
\hline\hline
$M_1$ & $\Gamma^{(1)}_{K\pi}$ & $\Gamma^{(1)}_{K\eta'}$ & $M_2$
& $\Gamma^{(2)}_{K\pi}$ & $\Gamma^{(2)}_{K\eta'}$ & $\chi^2$ \\
\hline
1.438 & 0.345 & 0.001 & --- & --- & --- & 86/45 \\
1.486 & 0.346 & 0.000 & 1.668 & 0.150 & 0.491 & 57/45 \\
\hline\hline
\end{tabular}
\caption[Parameters of resonances in $I=1/2~K\pi~S$-wave]
{\label{tab:kpiI1ResonancesParameters} Fitted parameters for the
s-channel resonances and the corresponding $\chi^2$ for two cases:
with one resonance (first line) and  with two resonances (second
line). Values for mass and width are in unit of GeV.}
\end{center}
\end{table}

It is natural that the fit with two s-channel resonances gives a
smaller $\chi^2$ value. But from
Fig.\ref{fig:kpiI1PhaseshiftAndAmplitude}, we see that both cases
with one or two s-channel resonances give quite good fit to the
data. For the case of two resonances, the second resonance is very
broad and has a mass above the upper energy limit (1.6 GeV) of the
data, and could be an effective tail of resonances above 1.6 GeV.
In both cases, there is only one s-channel resonance between the
$K\pi$ threshold and 1.6 GeV, corresponding to the PDG well
established $K^*(1430)$ resonance. The fitted mass and width for
the $K^*(1430)$ depend on whether we introduce one or two
s-channel resonances, with mass around $1438\sim 1486$ MeV and
width about 346 MeV, which is very close to the value (1450, 350)
MeV by Tornqvist and Roos \cite{Tornqvist1996:prl76} with a different
formalism.

For the $t$-channel $\rho$ meson exchange amplitude, we find a
pole at (0.45-0.48i) GeV. This is consistent with the conclusion
by Cherry and Pennington that there is no $\kappa(900)$, but a
very low mass $\kappa$ well below 825 MeV cannot be ruled out.

In summary, the $K\pi~I=3/2~S$-wave phase shift can be well
reproduced by the $t$-channel $\rho$ and $u$-channel $K^*$ meson
exchange while the $K\pi~I=1/2~S$-wave phase shift are dominated
by the s-channel $K_0^*(1430)$ resonance and the $t$-channel
$\rho$ exchange with a pole at ($450-480i$) MeV. The $\kappa(450)$
has a similar nature as
$\sigma(400)$\cite{LongLi2001:prd63,Zou1994:prd50}: both are
produced by the t-channel $\rho$ exchange and are very broad with
a width around 1 GeV.

%
\begin{acknowledgments}
This work was supported in part by the Major State Basic Research
Development Program (G20000774), CAS Knowledge Innovation Project
(KJCX2-SW-N02) and by National Natural Science Foundation of China
under Grant Nos. 19835010, 19905011 and 10175074.
\end{acknowledgments}
%
\newcommand{\mybib}[5]{{#1}, {#2} {\textbf{#3}}, {#4} {(#5)}}
%
\newcommand{\plb}{Phys. Letter. B}
\newcommand{\npa}{Nucl. Phys. A}
\newcommand{\npb}{Nucl. Phys. B}
\newcommand{\epja}{Eur. Phys. J A}
\newcommand{\epjc}{Eur. Phys. J. C}
\newcommand{\zpa}{Z. Phys. A}
\newcommand{\zpc}{Z. Phys. C}
\newcommand{\phr}{Phys. Rept.}
\newcommand{\ptp}{Prog. Theor. Phys.}
\newcommand{\etal}{\textit{et al.}}
%

\end{document}